\documentstyle[12pt,twoside,fleqn,espcrc1]{article}

\title{Regularization and renormalization in effective field
theories of the nucleon-nucleon interaction}

\author{D.~R.~Phillips, S.~R.~Beane and T.~D.~Cohen~\address{Department of
	Physics, University of Maryland,  
	College Park, MD, 20742-4111, USA}}

\begin{document}
% typeset front matter
\maketitle

\begin{abstract}
Some form of nonperturbative regularization is necessary if effective
field theory treatments of the $NN$ interaction are to yield finite
answers.  We discuss various regularization schemes used in the
literature. Two of these methods involve formally iterating the
divergent interaction and then regularizing and renormalizing the
resultant amplitude. Either a (sharp or smooth) cutoff can be
introduced, or dimensional regularization can be applied.  We show
that these two methods yield different results after
renormalization. Furthermore, if a cutoff is used, the $NN$ phase
shift data cannot be reproduced if the cutoff is taken to infinity.
We also argue that the assumptions which allow the use of dimensional
regularization in perturbative EFT calculations are violated in this
problem. Another possibility is to introduce a regulator into the
potential before iteration and then keep the cutoff parameter
finite. We argue that this does not lead to a
systematically-improvable $NN$ interaction.
\end{abstract}

\section{EFFECTIVE FIELD THEORIES OF THE NUCLEON-NUCLEON INTERACTION}

There exist many nucleon-nucleon potentials which reproduce phase
shifts and nuclear properties with remarkable
accuracy~\cite{NNpotnls}. Three fundamental features are shared by
these potential models: (i) pions dominate at long distances, (ii)
there is a source of intermediate-range attraction, and (iii) there is
a source of short-distance repulsion. Generally, distinct physical
mechanisms in these models account for the same feature of the nuclear
force. The agreement with experiment is maintained in spite of this
because of the large number of fit parameters.  It would be a
considerable advance in nuclear physics if a systematic approach to
the nucleon-nucleon potential were developed based solely on
symmetries and general physical principles.

One reason to hope that this can be achieved in nuclear physics is
provided by the pattern of chiral symmetry breaking in QCD. The fact
that chiral symmetry is spontaneously broken implies that the pion is
light and interacts weakly at low energies. The lightness of the pion
guarantees that it should play a fundamental role in nuclear
physics. The weakness of pion interactions at low energies allows pion
interactions with nucleons to be systematized using chiral
perturbation theory. This procedure has proved remarkably successful
in describing the interactions of pions with a single
nucleon~\cite{Ber95}.

Systematic approaches to the scattering of strongly interacting
particles, such as chiral perturbation theory, are based on the ideas
of effective field theory (EFT).  Effective field theory says that for
probes of a system at momentum $k \ll M$, details of the dynamics at
scale $M$ are unimportant. What is important at low energies is the
physics that can be captured in operators of increasing dimensionality
which take the form of a power-series in the quantity,
$k/M$~\cite{EFT}. 

In order to examine the application of EFT to nuclear physics,
consider $NN$ scattering in an $S$-wave channel at momentum scales $k
\ll m_\pi$.  The EFT at these scales involves only nucleons since the
pion is heavy relative to the energy scales under consideration and
may therefore be ``integrated out''. The effective Lagrangian then
consists of contact operators of increasing dimensionality constrained
by spin and isospin.  One might naively expect to be able to calculate
the $NN$ scattering amplitude directly from the effective Lagrangian
as a power series in $k/ {m_\pi}$:
\begin{equation}
T(k)=C + C_2 k^2 + C_4 k^4 + \ldots,
\label{eq:Texp}
\end{equation}
where with the prevailing prejudice of EFT we anticipate that the
dimensional coefficients $C_{2n}$ will be ``natural'', i.e. of order
$m_\pi^{-2-2n}$. We know that in the ${}^1S_0$ and ${}^3S_1-{}^3D_1$
channels there are, respectively, a quasi-bound state and a bound
state at low energies. The power series expansion (\ref{eq:Texp}) with
natural coefficients can only be correct if these bound states are at
``natural'' energies $k^2 \sim m_\pi^2$. However, in these channels
the bound states occur at unnaturally low energies. Therefore, the
coefficients in the expansion must be unnatural: they are fixed by the
pole positions of the low-lying bound states rather than the scale of
the physics that has been integrated out.  This limits the usefulness
of an expansion in the amplitude to very low energies.  On the other
hand, if, following Weinberg's suggestion~\cite{Wein}, one makes an
EFT expansion of the potential in $S$-wave channels and iterates it
via the Lippmann-Schwinger equation:
\begin{eqnarray}
V(p',p')&=&C + C_2 (p^2 + p'^2)+ C_4
(p^4 + p'^4) + C_4' p^2 p'^2 + \ldots
\label{eq:Vexp};\\
T(p',p;k)&=&V(p',p) + M\int \frac{d^3p''}{(2 \pi)^3} \, V(p',p'') 
\frac{1}{k^2- {p''^2}+i\epsilon} T(p'',p;k),
\label{eq:LSE}
\end{eqnarray}
one may hope to generate (quasi-)bound states at the appropriate
energies while maintaining natural coefficients in the
potential. There will, of course, be unknown coefficients in the 
potential, but these can be fit to experimental
data~\cite{Or96,Ka96,Sc97,Le97} as in ordinary chiral perturbation
theory.

At face value this procedure appears promising. For momenta $p,p' \sim
k \ll m_\pi$ the potential (\ref{eq:Vexp}) may be approximated by its
first few terms. More generally, beginning from the Lagrangian of
chiral perturbation theory, Weinberg has shown how to construct a
nucleon-nucleon potential of definite chiral
order~\cite{Wein,Or96}. In this potential pions are explicit degrees
of freedom, while all other particles are ``integrated out'',
i.e. their effects appear in the Lagrangian via a derivative
expansion, \'a la Eq.~(\ref{eq:Vexp}).  However, in iterating this
potential via Eq.~(\ref{eq:LSE}) a number of issues arise which are
absent in standard EFT treatments.

The main difficulty concerns nonperturbative regularization and
renormalization. This is required when iterating the potential
(\ref{eq:Vexp}) to all orders using Eq.~(\ref{eq:LSE}), since the hard
asymptotic behavior inherent to the momentum expansion necessarily
introduces divergences. The existence of a procedure to regularize
these divergences and renormalize in a sensible way provides a
non-trivial condition on the existence of an EFT. 

\section{THE $NN$ AMPLITUDE AT SECOND ORDER IN A SIMPLE EFFECTIVE FIELD THEORY}

Consider an effective field theory for the nucleon-nucleon interaction
in which all exchanged particles are integrated out. At second order
in this EFT the nucleon-nucleon potential in the ${}^1S_0$ channel may
be written in momentum space as:
\begin{equation}
V(p',p)=C + C_2 (p^2 + p'^2).
\end{equation}
Iterating using Eq.~(\ref{eq:LSE}), we get the following
result for the on-shell T-matrix~\cite{Ph97}:
\begin{equation}
\frac {1}{T(k)}= \frac{(C_2 I_3 - 1)^2}{C + C_2^2 I_5 + k^2 C_2 (2 - C_2 I_3)}
- I_1,
\label{eq:ampres}
\end{equation}
\begin{equation}
I_5=-M \int \frac{d^3p''}{(2 \pi)^3} p''^2; \quad
I_3=-M \int \frac{d^3p''}{(2 \pi)^3}; \quad
I_1=M \int \frac{d^3p''}{(2 \pi)^3} \frac{1}{k^2 - p''^2 + i\eta},
\end{equation}
where $k$ is the on-shell momentum.  The integrals $I_1$, $I_3$, and
$I_5$ all contain power-law divergences. Note that if $C_2=0$ this
reduces to the expression derived by Weinberg~\cite{Wein} for the
case where the potential is just $V(p',p)=C$.

We renormalize the amplitude (\ref{eq:ampres}) using the experimental
values of the scattering length, $a$, and effective range, $r_e$. In
other words, we fix $C$ and $C_2$ by demanding that
\begin{equation}
\frac{1}{T(k)}=-\frac{M}{4 \pi}\left(-\frac{1}{a} + \frac{1}{2} r_e k^2 
+ O(k^4) - i k \right).
\end{equation}
The divergent integrals $I_1$, $I_3$, and $I_5$ may be regularized in
two different ways: either a (sharp or smooth) cutoff can be
introduced, or dimensional regularization (DR) can be applied.  In DR
the power-law divergent pieces of $I_1$, $I_3$ and $I_5$ all vanish by
prescription. The renormalization then yields:
\begin{equation}
\frac{1}{T_{\rm DR}(k)}=-\frac{M}{4 \pi}
\left(\frac{1}{-a - \frac{1}{2} a^2 r_e k^2} - ik \right),
\label{eq:DRamp}
\end{equation} 
for all $a$ and $r_e$, as found by Kaplan {\it et al.}~\cite{Ka96}. By
contrast, if a cutoff is used then, once renormalization is
performed and the cutoff is taken to infinity we get~\cite{Ph97}
\begin{equation}
\frac{1}{T_{\rm cutoff}(k)}=-\frac{M}{4 \pi}\left(-\frac{1}{a} + \frac{1}{2} 
r_e k^2 - i k \right),
\label{eq:cutoffamp}
\end{equation}
{\it but only if $r_e \leq 0$}. If $r_e > 0$ the renormalization
cannot be performed in the limit that the cutoff is taken to infinity
if the coefficients $C$ and $C_2$ are real~\cite{Ph97}. This is
qualitatively different to the result found using dimensional
regularization. Even in the case where $r_e < 0$ the two
regularization schemes lead to amplitudes with different dependence on
$k$. In this nonperturbative problem the two methods of
regularization are inequivalent.

Recall that the equivalence of DR and other forms of regularization
has only been proven for perturbative calculations. In such cases all
power-law divergences can be absorbed into counterterms, so DR's
neglect of these divergences does not lead to any difference in the
renormalized amplitude. However, here the renormalization conditions
are highly nonlinear and the two forms of regularization are no
longer equivalent.

The result obtained here using cutoff regularization is in accord with
that found if the regularization is performed before iteration, as it
was in Refs.~\cite{Or96,Sc97,Le97}. In that case, a
corollary~\cite{PC97} of a theorem due to Wigner~\cite{Wi55}
states that if a potential goes to zero beyond some range $R$ then
\begin{equation}
r_e \leq 2 \left(R - \frac{R^2}{a} + \frac{R^3}{3 a^2}\right).
\label{eq:WB}
\end{equation}
In particular, this implies that there is a minimal range for the
regulated delta function below which the effective range cannot be
reproduced. Furthermore, a minimal range for the regulated contact
interactions exists even if pions are explicitly included in the
calculation~\cite{Sc97}.  These results hold whenever
the EFT expansion of the potential (\ref{eq:Vexp})
is truncated at some finite order.

In fact, they merely reflect a very basic fact of nuclear physics: the
range of the potential used to model the nuclear force is
important. To any finite order in momentum the potential
(\ref{eq:Vexp}) is a zero-range potential in coordinate space. By
regularizing this potential we give the interaction some finite
range. However, if this range is taken to zero (or, equivalently, the
momentum-space cutoff is taken to infinity), Eq.~(\ref{eq:WB}) shows
that it is impossible to reproduce the observed energy-dependence
of the $S$-wave $NN$ phase shifts.

Why then, is the renormalized amplitude obtained using DR not subject
to the bound (\ref{eq:WB})? In fact the power-law divergences which DR
discards carry crucial information on the range of the
interaction. DR's removal of these power-law divergences is justified
in perturbative EFT calculations, since such calculations are based on
the assumption that details of the short-distance physics, such as
power-law divergences, are unimportant. However, in this
nonperturbative problem physics at short, but finite, distances
carries important additional information which affects the
renormalized amplitude and so should not be discarded.

\section{CUTOFF EFFECTIVE FIELD THEORY FOR THE POTENTIAL}

Thus a cutoff cannot be introduced and then taken to infinity if one
wishes to reproduce the energy-dependence of the $NN$ phase shifts,
and the assumptions underpinning the use of DR in EFT calculations are
not satisfied in this problem.  

An alternative is to introduce a regulator into the potential before
iteration, and then keep its mass finite.  In such an approach
renormalization is performed by adjusting the potential to fit the
$NN$ scattering data. This can be done for several different values of
the regulator parameter. (For actual examples of this being done in
the $NN$ problem see~\cite{Or96,Sc97,Le97}.) In this section we
explain why this approach is not systematic.

Suppose that a new cutoff scale $\beta < m_\pi$ is introduced.  This
can be implemented by writing the potential in this cutoff field
theory for $NN$ scattering in the ${}^1S_0$ channel as
\begin{equation}
V(p',p)=[C + C_2 (p^2 + p'^2) + \ldots] \theta(\beta - p)
\theta(\beta - p').
\label{eq:Veffexp}
\end{equation}
The coefficients $C$, $C_2$, etc.  will all be functions of $\beta$.
This effective potential is to be inserted into Eq.~(\ref{eq:LSE}) and
the equation solved for different values of $\beta$ with the
coefficients $C(\beta)$, $C_2(\beta)$, etc. fit to low-energy
scattering data. Of course, as $\beta$ is varied
the coefficients will change significantly.

If power counting for the potential were correct, and the expansion
(\ref{eq:Veffexp}) was only used within its regime of validity, then
the terms in the potential would get systematically smaller as the
``order'' is increased.  However, in fact this does not happen---at
the upper end of the momentum domain all terms contribute equally to
the potential.  One simple way to see this is by looking at what
happens if a different form of regulator is chosen.

Of course, the choice of a theta function to regulate the
momentum-space integrals is entirely arbitrary. One could just as well
choose a potential of the form
\begin{equation}
V(p',p)=[\tilde{C} + \tilde{C_2} (p^2 + p'^2) + \ldots]
g(p^2/\beta^2,p'^2/\beta^2),
\label{eq:Veffexp2}
\end{equation}
where $g$ is some suitably well-behaved function. The effective
potential should be essentially unaltered by this change in the form
of the cutoff. However, this implies that the ratios
$\tilde{C_n}/\tilde{C}$ differ from the ratios $C_n/C$ by terms of
order $1/\beta^n$.  Therefore for a generic cutoff function $g$ the
ratio $C_n/C$ must be of order $1/\beta^n$.  If this is the case, and
the effective potential is to be used in a momentum regime which
extends up to $\beta$, then at the upper end of this momentum regime
all terms in the expansion are equally important. Therefore one should
not truncate Eqs.~(\ref{eq:Veffexp}) or (\ref{eq:Veffexp2}) at some
finite order in $p$ and $p'$. This is not surprising: if a finite
momentum-space cutoff is used, the resulting potential contains all
orders in momentum, and so is sensitive to physics at arbitrarily
short distances.

Therefore, none of the regularization schemes discussed here give a
systematic EFT $NN$ potential. A paper discussing these
issues in more detail is in preparation~\cite{Be97}.

\section*{Acknowledgments}
We thank M.~C.~Birse, J.~A.~McGovern, E.~F.~Redish, U.~van Kolck,
S.~J. Wallace, and S. Weinberg for useful and enjoyable
discussions on the topics discussed in this paper. This research was
supported in part by the U. S. Department of Energy, Nuclear Physics
Division (grant DE-FG02-93ER-40762).

\end{document}